\documentclass[preprint,showpacs,preprintnumbers,amsmath,amssymb,nofootinbib]{revtex4}

\usepackage{etex}
\usepackage{graphicx}
\usepackage{dcolumn}
\usepackage{bm}
\usepackage{amssymb,amsthm,amscd,amsbsy,array}\usepackage{amsmath,dsfont}

\usepackage{graphics,xcolor}

\usepackage{color}
\newcommand{\red}{\textcolor{red}}
\newcommand{\blue}{\textcolor{blue}}

\newcommand{\PT}{{P\"oschl - Teller\;}}

\newcommand{\GW}{{gravitational wave\;}}
\newcommand{\DM}{{Displacement Memory\;}}

\newcommand{\half}{{\scriptstyle{\frac{1}{2}}}}
\def\2{{\half}}
\newcommand{\const}{\mathop{\rm const}\nolimits}

\def\p{{\partial}}
\def\bk{{\bf k}}

\def\bp{{\bm{p}}}

\def\bp{{\bm{p}}}

\def\bX{{\bm{X}}}

\def\beqa{\begin{eqnarray}}
\def\eeqa{\end{eqnarray}}

\def\barray{\left(\begin{array}}
\def\earray{\end{array}\right)}
\def\barraynb{\begin{array}}
\def\earraynb{\end{array}}

\def\smallover#1/#2{\hbox{$\textstyle\frac{#1}{#2}$}} %

\newcommand{\cA}{{\mathcal{A}}}

\def\smallcirc{{\raise 0.5pt \hbox{$\scriptstyle\circ$}}}
\def\aand{{\quad\text{\small and}\quad}}

\def\with{{\quad\text{\small with}\quad}}

\def\ie{{\;\text{\small i.e.}\;}}
\def\ie,{{\;\text{\small i.e.,}\;}}

\def\GW{{gravitational wave\,}}

\def\VM{{Velocity Effect\,}}
\def\DM{{Displacement Effect\,}}

\def\benu{\begin{enumerate}}
\def\eenu{\end{enumerate}}
\def\bitem{\begin{itemize}}
\def\eitem{\end{itemize}}

\def\besub{\begin{subequations}}
\def\esub{\end{subequations}}

\def\?{{\,\gb{\fbox{\texttt{??}}\;}}\,}

\def\StL{{Sturm-Liouville\,}}
\def\cI{{${\cal I}$}}
\def\cI{{{\cal I}}}

\usepackage[colorlinks=true, pdfstartview=FitV, linkcolor=blue, citecolor=blue, urlcolor=blue]{hyperref} 

\newcommand{\purple}{\textcolor[rgb]{0.5,0.0,0.5}}
\newcommand{\gb}{\quad\colorbox{green}}

\newcommand{\dgreen}{\textcolor[rgb]{0,0.5,0}}

\newenvironment{redtext}{\color{red}}
{\ignorespacesafterend}
\newenvironment{bluetext}{\color{blue}}{\ignorespacesafterend}
\newenvironment{greentext}{\color{green}}{\ignorespacesafterend}
\newenvironment{magentatext}{\color{magenta}}{\ignorespacesafterend}
\newenvironment{cyantext}{\color{cyan}}{\ignorespacesafterend}
\newenvironment{orangetext}{\color{orange}}
{\ignorespacesafterend}

\newcommand{\bmagenta}{\begin{magentatext}}
\newcommand{\emagenta}{\end{magentatext}}
\newcommand{\bcyan}{\begin{cyantext}}
\newcommand{\ecyan}{\end{cyantext}}
\newcommand{\bblue}{\begin{bluetext}}
\newcommand{\eblue}{\end{bluetext}}
\newcommand{\bred}{\begin{redtext}}
\newcommand{\ered}{\end{redtext}}

\newcommand{\bgreen}{\begin{greentext}}
\newcommand{\egreen}{\end{greentext}}
\newcommand{\borange}{\begin{orangetext}}
\newcommand{\eorange}{\end{orangetext}}

\numberwithin{equation}{section}

\let\ssection=\section
\renewcommand{\section}{\setcounter{equation}{0}\ssection}
\newcommand{\beq}{\begin{equation}}
\newcommand{\eeq}{\end{equation}}
\newcommand{\bec}{\begin{center}}
\newcommand{\ec}{\end{center}}

\usepackage{amsthm}

{Corollary}[section]

\begin{document}

\preprint{arXiv:2603.15442v3}

\title{Approximate Models for Gravitational Memory}

\author{
Q-L Zhao $^{1}$\footnote{mailto:zhaoqliang@ucas.ac.cn},
P.-M. Zhang$^{2,3}$\footnote{Corresponding author mailto:zhangpm5@mail.sysu.edu.cn},
M. Elbistan$^{3}$\footnote{mailto:mahmut.elbistan@bilgi.edu.tr}
and
P. A. Horv\'athy$^{3,4}$\footnote{mailto:horvathy@univ-tours.fr}
}

\affiliation{
${}^{1}$ School of Fundamental Physics and Mathematical Sciences,
		Hangzhou Institute for Advanced Study, UCAS, Hangzhou 310024, China	
\\
$^2$ School of Physics and Astronomy, Sun Yat-sen University, Zhuhai 519082, (China)
\\
${}^3$ Department of Energy Systems Engineering, Istanbul Bilgi University, 34060, Eyupsultan, Istanbul, (Turkey)
\\
${}^{4}$ Institut Denis-Poisson CNRS/UMR 7013 - Universit\'e de Tours - Universit\'e d'Orl\'eans Parc de Grammont, 37200; Tours, (France).
}
\date{\today}
\begin{abstract}
The large-distance approximation
 of a sandwich gravitational wave by a continuous but not necessarily smooth profile provides us with an  approximate analytic description of  particle motion in a gravitational wave as spelled out for the P\"oschl-Teller profile.  Displacement Memory is obtained by fine-tuning the amplitude. The role of the 2nd solution of the Sturm-Liouville equation is highlighted. Similar results hold for a Gaussian and simple square profiles. Our approximate models are consistent with  Carroll symmetry.%
\end{abstract}


\maketitle

\tableofcontents

\section{Introduction: }\label{Intro}

An early proposal to detect
 gravitational waves was  to observe the displacement of  particles hit by a burst of sandwich waves \cite{BraTho}, called the \emph{Memory Effect} (ME). Progress using space-based detectors (as LISA) might lead to test the proposal.

Initial study \cite{Ehlers} hinted at the
\emph{\VM} (VM):  particles initially at rest would fly off with \emph{constant non-zero velocity} after the wave has left. Zel'dovich and Polnarev \cite{ZelPol} argued instead in favour of the \emph{Displacement Effect} (DM), suggesting  that the particles would merely be displaced.
Their statement could be tested by studying 
 various wave profiles including a Gaussian \cite{LongMemory,EZHRev} or \PT,  their derivatives \cite{PTeller, Chakra, DM-1,DM-2}, or the Scarf   \cite{Scarf,Sila-PLB}  potentials, and for a simple square profile \cite{Kar3,Benin}. These results confirm the  Zel'dovich - Polnarev statement  \cite{ZelPol} \emph{provided the wave amplitude takes some ``magic'' values} \cite{DM-1,DM-2,Jibril,Benin}.
\goodbreak

\section{Memory effect in a plane \GW} \label{MemoGWSec}

We consider a linearly polarised vacuum wave with Brinkmann  metric,
\beq
g_{\mu\nu} dX^{\mu} dX^{\nu} = d\bX^2+2dUdV + \cA(U)\Big((X^{+})^2-(X^{-})^2\Big)
\,dU^2\,,
\label{Bplanewave}
\eeq
where the
 $\bX = (X^+,X^-)$ are coordinates of the transverse plane with  flat Euclidean metric $d\bX^2=\delta_{ij}\,dX^idX^j$; $U$ and $V$ are light-cone coordinates.
 $\cA(U)$ is  the profile of the wave.
 The relative minus follows from the vacuum Einstein equations \cite{exactsol}.
 The eqns of motion of a particle we took massless for simplicity are,
\begin{subequations}
\begin{align}
&{\dfrac {d^2\!X^+}{dU^2} - \cA(U) X^+ = 0\,,} \label{geoX1}
\\[6pt]
&{\dfrac {d^2\! X^-}{dU^2} + \cA(U)  X^- = 0\,,}
\label{geoX2}
\\[8pt]
&
\dfrac {d^2\!V}{dU^2}
+\frac{1}{2}\dfrac{d\cA}{dU}\Big((X^+)^2-(X^-)^2\Big)
 +
2\cA\Big(X^+\frac{dX^+}{dU}-X^-\dfrac{dX^-}{dU}\Big)=0\,.
\label{geoVfly}
\end{align}
\label{Bgeoeqn2}
\end{subequations}
%
Eqn. \eqref{geoVfly}  corresponds to the null lift of $\bX$ and thus follows from \eqref{geoX1} - \eqref{geoX2}, which describe motion in  a  harmonic force. For $\cA >0$ the $X^+$ sector is repulsive and that of $X^-$ is attractive.

 Below study the  attractive dynamics of $X \equiv X^-$ in  $D=1$-dimension, eqn. \eqref{geoX2}, spelling out our observations  for the analytically solvable \PT profile \cite{PTeller,Chakra,DM-1}.
The initial conditions for a  particle at rest before the wave arrives are,
\beq
X(-\infty)= X_{0}=\const
\qquad
\frac{dX}{dU}(-\infty) = 0\,.
\label{initcond}
\eeq
The \DM\! (DM) arises when we have, in addition,
\beq
X(+\infty)= X_{\infty}=\const,
\qquad
\frac{dX}{dU}(+\infty) = 0\,.
\label{DMboundcond}
\eeq

\section{An approximate profile for \PT }\label{ApproxToy}

A  simple example considered before in \cite{Chakra,DM-1,DM-2} is given by the \PT potential \cite{PTeller},
\beq
\cA^{PT}(U) = \dfrac{m(m+1)}{\cosh^2 U}\,,
\label{PTellerPot}
\eeq
where $m$ is a real constant, related to the wave amplitude, $k^2=m(m+1)$.
The \StL eqn \eqref{geoX2}, which is now
\begin{equation}
\dfrac{d^{2}X}{dU^2}+
\frac{m\left(m+1\right)}{\cosh^{2}U}X=0\,,
\label{PTtraj}
\end{equation}%
satisfies the DM condition \eqref{DMboundcond}
 when $m$ is a positive integer \cite{DM-1,Chakra}.
The solutions of \eqref{PTtraj} are then proportional to  Legendre polynomials\footnote{No summation understood.},
\beq
X(U) \equiv  X_m(U) = (-1)^m\,P_m(\tanh U)\,X_0,
\qquad
m = 1,\, 2,\, \dots\,
\label{PT-XmU}
\eeq
The trajectories are  composed of $m$ half-waves \cite{DM-1,DM-2}~\footnote{When the amplitude
is negative, the  solutions of \eqref{PTtraj} are  Legendre functions and the required boundary conditions can not be satisfied. We assume hence that we are in the attractive sector, $m > 0$.}. Note for further reference that the  properties of Legendre polynomials imply that the  trajectories \eqref{PT-XmU} satisfy
\beq
X_m^{out} \equiv X_m(+\infty) = (-1)^m X_m(-\infty) \equiv X_m^{in}\,.
\label{Xminuplusinf}
\eeq

\begin{figure}[h]
\includegraphics[scale=.85]{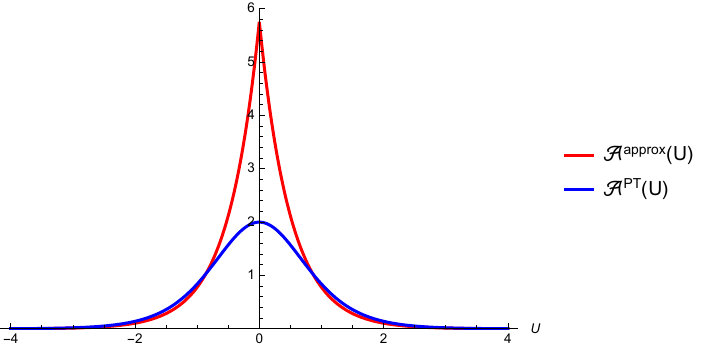}
\vskip-3mm\caption{
 \textit{\small For large $|U|$, the \red{\bf approximate profile} \eqref{extoy} with amplitude $k\approx 2.4$ is close to that of \blue{\PT\!}  $\cA^{PT}$ with $m=1$, in \eqref{PTellerPot}.
However $\cA^{approx}$ in \eqref{extoy} has a peak at $U=0$ and the two profiles differ substantially. }
\label{exp-PT} }
\end{figure}

An approximate ``toy'' model \cite{ZZH22}  is obtained as follows. For large $U$, we have
$\cosh^{-2}(U)\approx 4e^{-2|U|}$,
and we propose to approximate the \PT profile  \eqref{PTellerPot} for $m=1$ by
\beq
\cA^{approx}(U) = k^2e^{-2|U|}\,
\label{extoy}
\eeq
where the parameter $k \approx 2.4$ is obtained by fine-tuning for DM. For large $|U|$, the profile is close to that of PT in \eqref{PTellerPot}, as seen in FIG.\ref{exp-PT}. The approximation breaks manifestly down in the neighbourhood of the origin.

\begin{figure}[h]
\includegraphics[scale=.29]{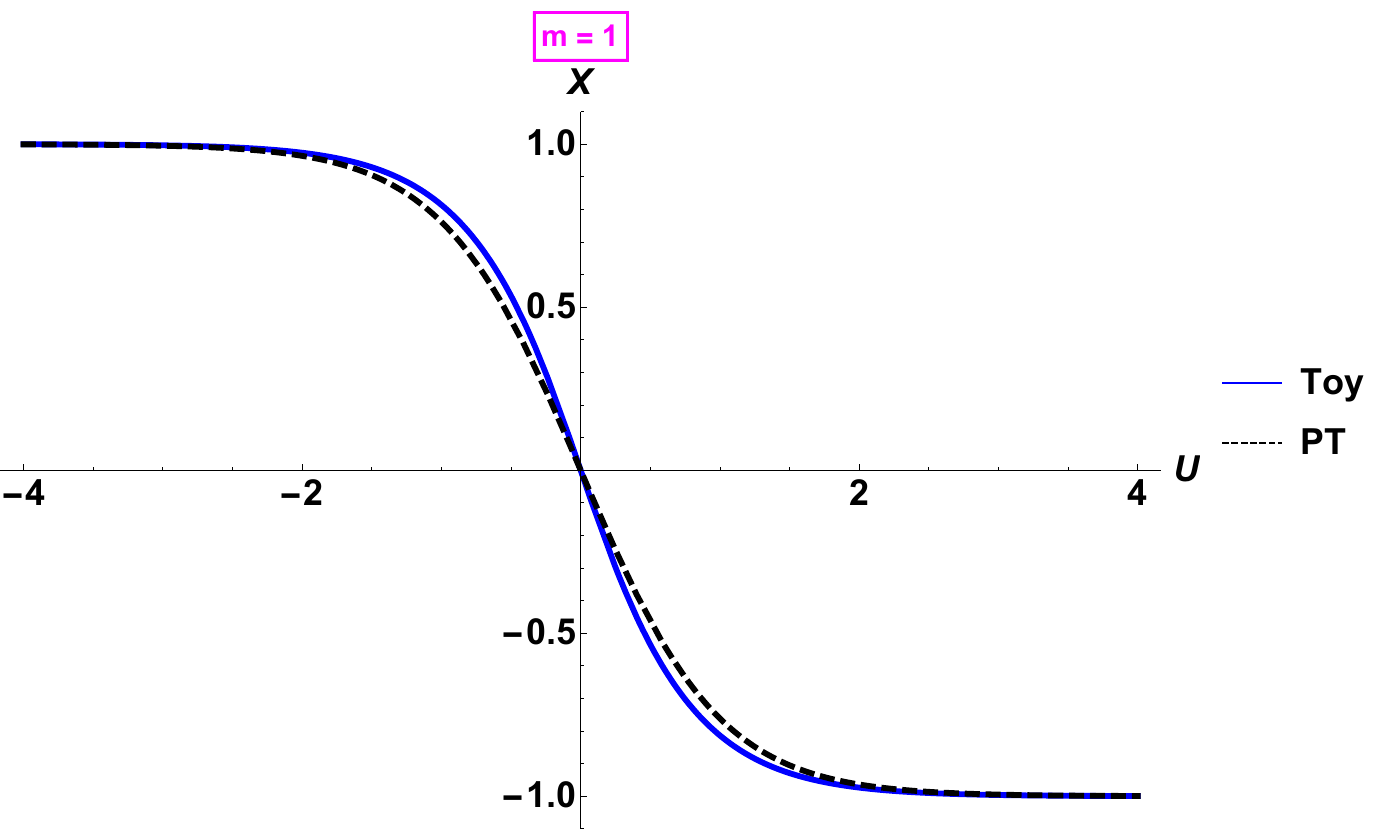}\quad\,
\includegraphics[scale=.31]{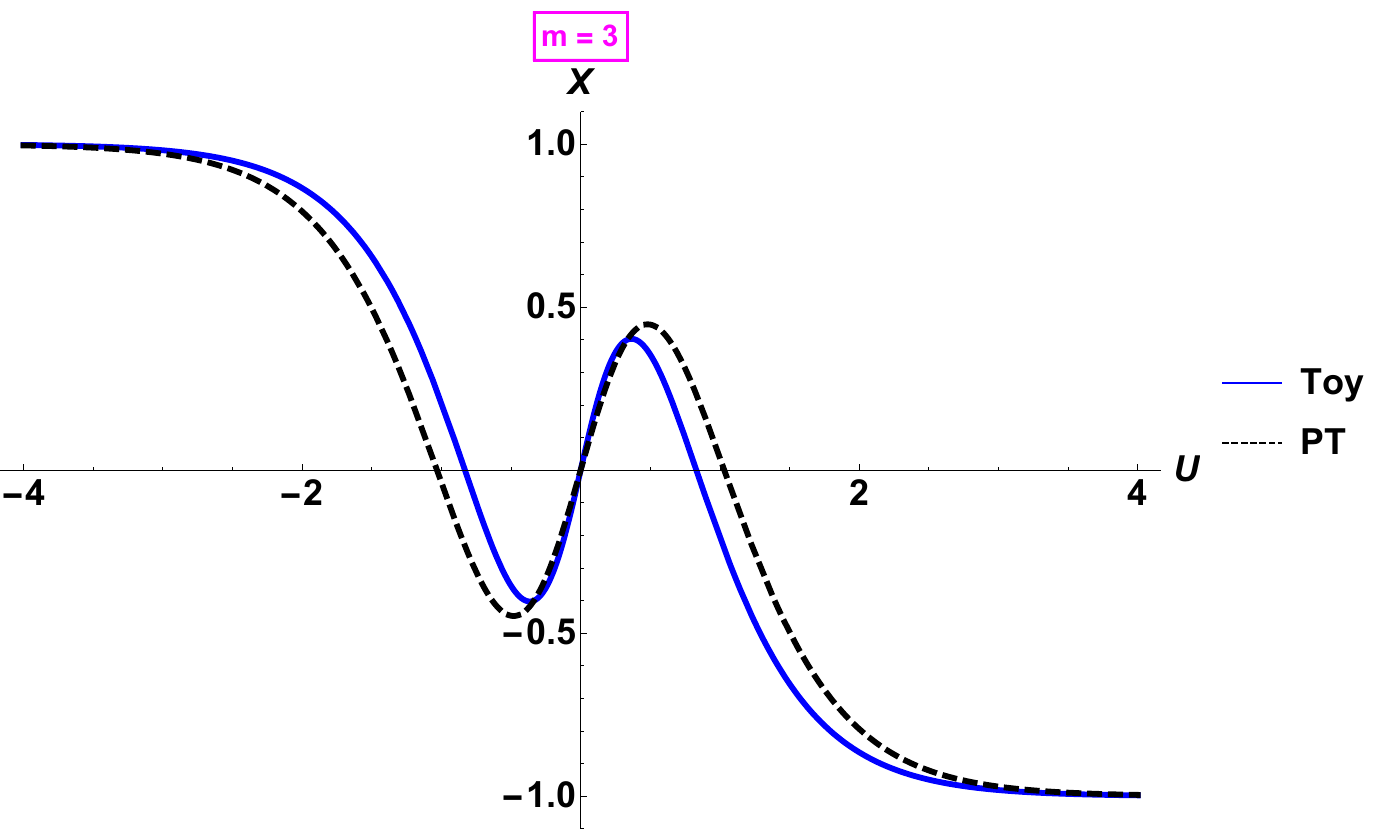}
\vskip-1mm
\hskip-13mm (a) \hskip68mm (b)
\vskip-3mm \caption{\textit{\small
DM  trajectories (dashed) for the {\bf toy model} \eqref{extoy}
 with $X_{out}= -X_{in}$ are obtained
  when their amplitude is a zero of the Bessel function $J_0$. 
 They approximate  those of \blue{\bf \PT} in \eqref{PT-XmU} with {\bf odd} half-wave number $m=2\ell+1$.}
\label{modd} }
\end{figure}

The approximate model has exact geodesics, given by a combination of Bessel functions of order 0 of the first and of the second kind  \cite{ZZH22},
\beq
X(U) = \left\{\barraynb{clccl}
\alpha_1\,J_0\big(ke^{-U}\big) &+ &\beta_1\,Y_0\big(ke^{-U}\big) &\;\text{in}\;\;&\cI_{+}
\\[4pt]
\alpha_2\,J_0\big(ke^{U}\big) &+ &\beta_2\,Y_0\big(ke^{U}\big) &\;\text{in}\;\;& \,\cI_{-}
\earraynb\right.\;
\,,
\label{J0Y0}
\eeq
where 
$
\cI_{-} = \big\{U <0\big\}
$ and $
\cI_{+} = \big\{U > 0\big\}\,.
$
The DM boundary conditions $X^{\prime}(\pm\infty)=0$ require to set  $\beta_1=\beta_2= 0$, leaving us with
\beq
X(U) = \left\{\barraynb{llcll}
X_{+}(U)&=& \alpha_1J_0\big(ke^{-U}\big)  &\text{in}\;& \cI_{+}
\\[4pt]
X_{-}(U)&=&\alpha_2\,J_0\big(ke^{U}\big)  &\text{in}\;& \cI_{-}
\earraynb\right.
\,,
\label{regBessel}
\eeq
where $\alpha_1$ and $\alpha_2=X_0$ are arbitrary constants.
Physically admissible trajectories are obtained when the left and right solutions match {smoothly}. For this we should have, first of all,
\beq
X_{-}(0)=\alpha_2 J_0(k)
=
X_{+}(0)=\alpha_1 J_0(k)
\label{fitX}
\eeq
for which we have two possibilities~:
\benu
\item
When  
$ 
J_0(k)=0$ which happens for amplitudes $k_m = 2.4,\, 5.5, \, 8.7, \, \dots
$ with odd wave number $m = 2\ell+1$.
Then the trajectory is continuous for arbitrary
  $\alpha_1,\, \alpha_2$~: the two branches are joined at the origin, 
\beq
  X_-(0)=X_-(0)=0\,.
\label{oddcondition}
 \eeq  
However the slopes must be equal also,
\beq
X_{-}'(0)=X_{+}'(0)\,,
\label{fitslope}
\eeq
which then requires
$\alpha_1=-\alpha_2\equiv -\alpha$.
The trajectory is obtained by glueing the $U < 0$ and $U > 0$ branches antisymmetrically,
\beq
X(-U) = - X(U)\,
\label{asymX}
\eeq
which is consistent with \eqref{oddcondition}.   
 Thus we end up with
\beq
X^{odd}(U) = \left\{\barraynb{llcll}
X_{+}(U)&=& - \alpha\, J_0\big(ke^{-U}\big)  &\;\text{in}\;& \cI_{+}
\\[3pt]
X_{-}(U)&=&\,\alpha \,J_0\big(ke^{U}\big)  &\;\text{in}\;& \cI_{-}
\earraynb\right.
\,.
\label{DModd}
\eeq
Despite the lack of smoothness of the profile,
the approximate trajectories \eqref{DModd} have surprisingly similar shape to the
usual  \PT  ones in  \eqref{PT-XmU} with
{odd} half-wave number, see FIG.\ref{modd}. 

\begin{figure}[h]
\includegraphics[scale=.3]{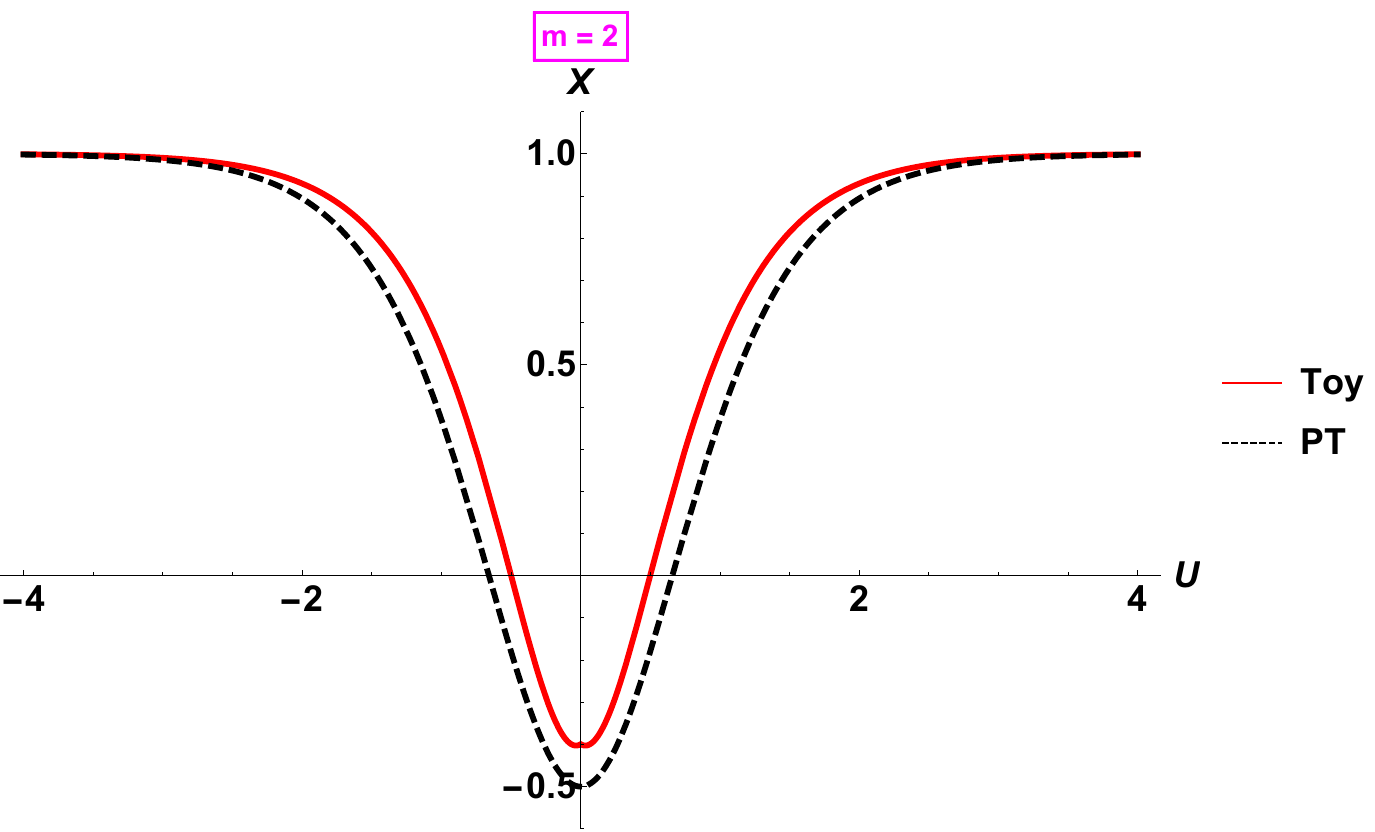}\;\;\;\;
\includegraphics[scale=.3]{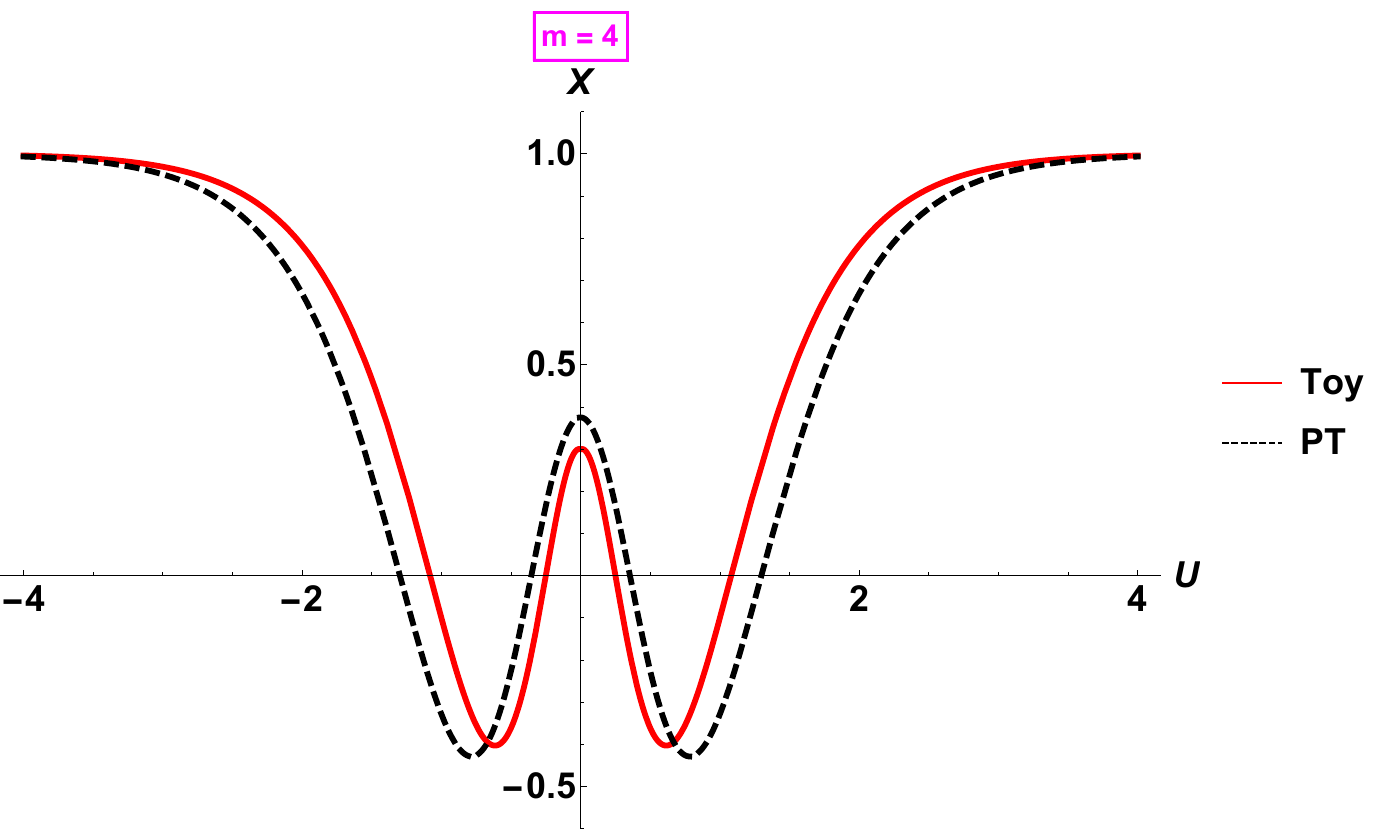}
\vskip-2mm
\hskip-10.5mm (a) \hskip69mm (b)
\vskip-3mm \caption{\textit{\small
DM toy trajectories (dashed) are obtained  when the amplitude is a zero of the Bessel function $J_1$.
The Wavezone contains an {\bf even} number $m=2\ell$ of symmetrically positioned half-waves which approximate those for \red{\bf \PT\!}.
 }
\label{meven} }
\end{figure}

\item
There is yet another possibility with $J_0(k)\neq0$~: when $\alpha_1=\alpha_2\equiv\alpha $ then the first fitting condition $X_{-}(0)=X_{+}(0)=J_{0}(k)$ is satisfied for \emph{any} amplitude $k$.
However the trajectory must be also smooth,
\beq
X_{-}'(k)=-(\alpha k) J_1(k)  =
X_{+}'(k)=  +  (\alpha k)J_1(k)\,,
\label{Xprimes}
\eeq
upon using $J_0'=-J_1$. Thus $k$ must be a root of $J_1$,
$ 
J_1(k)=0\,,\ie, \;
k = k_{(2\ell)} =  3.8,\, 7.0,\, 10.1,  \, \dots
$ 
\eenu
In conclusion, the DM trajectory with $\alpha=X_0$,
\beq
X^{even}(U) = J_0\big(k_{(2\ell)}e^{-|U|}\big)\,X_0 \,,
\label{eventraj}
\eeq
is a good approximation of the  \PT case \eqref{PT-XmU} for \emph{even} half-wave number $m=2\ell$. It
is obtained by gluing together the negative and positive $U$-branches \emph{symmetrically},
\beq
X(-U) =  X(U)\,,
\label{symX}
\eeq
yielding ``DM trajectories with no displacement'', shown in FIG.\ref{meven}.

\begin{figure}[h]
\includegraphics[scale=.73]{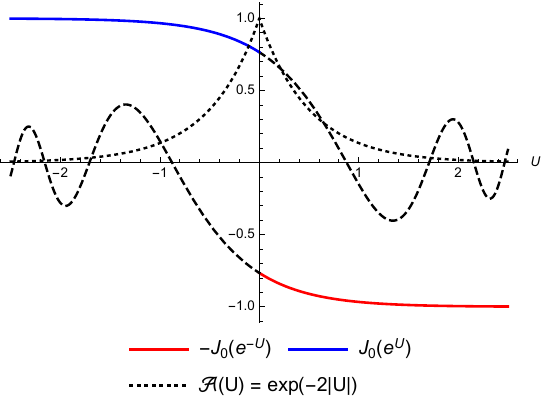}
\qquad\,
\includegraphics[scale=.73]{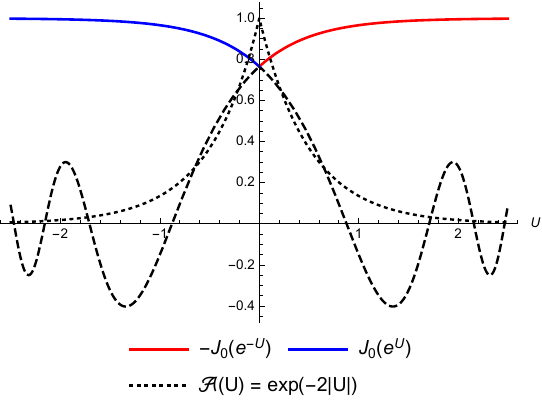}
\\
\vskip-3mm\hskip-3mm
(a) \hskip72mm (b) \vskip-3mm\caption{\textit{\small For amplitude  $k\neq k_{krit}$
 the positive and negative $U$-branches do not match smoothly~:
(a)  for  antisymmetric fitting  $X_{-}(0)\neq X_{+}(0)$  (b) for  symmetric fitting $X_{-}'(0)\neq X_{+}'(0)$.
}
\label{discont} }
\end{figure}

\begin{figure}[h]
\hskip-4mm
\includegraphics[scale=.32]{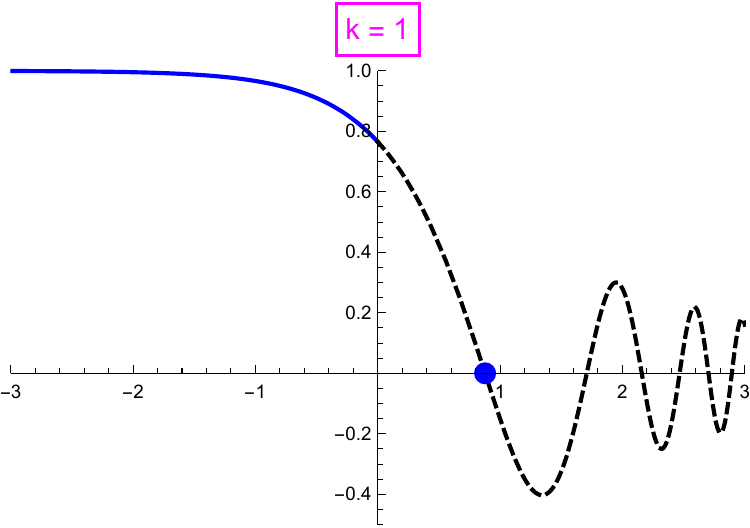}
\includegraphics[scale=.32]{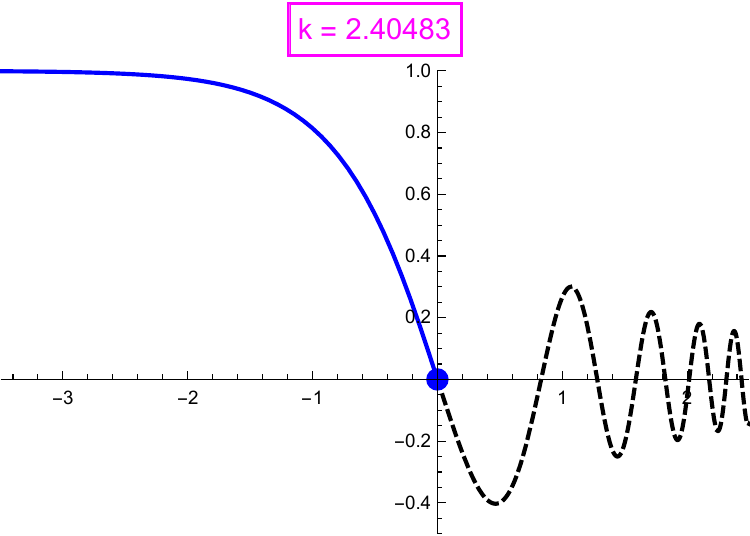}
\includegraphics[scale=.32]{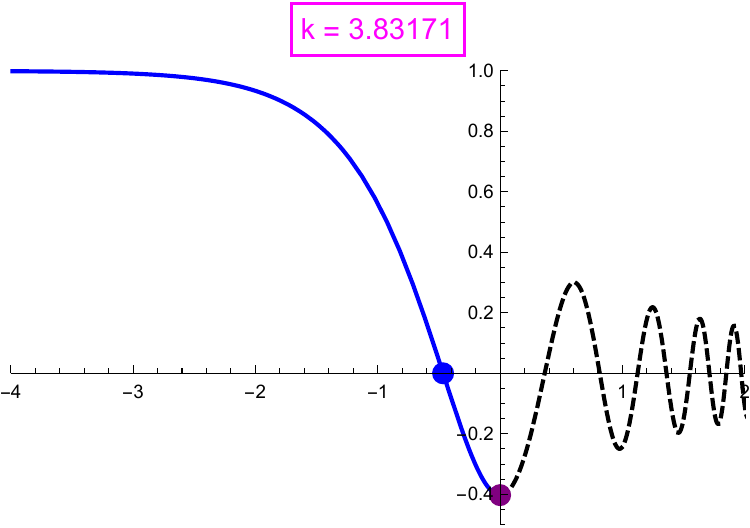}
\includegraphics[scale=.32]{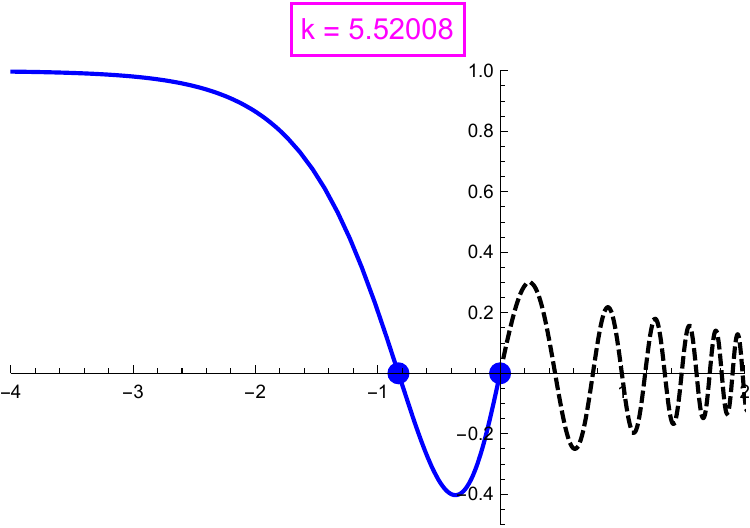}

\hskip2.5mm
(a)\hskip40.5mm (b)\hskip40mm (c) \hskip36mm (d)
\vskip-3mm\caption{\textit{\small Increasing the amplitude $k$ shifts the trajectory along $U$ leftwards.
DM is obtained when the far-most-left \blue{\bf zero} (in \blue{\bf blue}) or the far-most-left \purple{\bf bottom} (in \purple{\bf purple}) enters into the physical domain by reaching the $U=0$ axis.
}
\label{Zhaoshift} }
\end{figure}

An intuitive insight of how this comes about
 is gained by looking at  FIGs.\ref{discont}-\ref{Zhaoshift}.
%
 Consider first the antisymmetric fitting.
In  $\cI_{-}$ the trajectory
 has  $U\to 0^-$ limit $X^{-}(0)=\alpha J_0(k)$. The $U\to 0^+$ limit in $\cI_{+}$
 is instead  $-\alpha J_0(k)=-X^{-}(0)$, shown in FIG.\ref{discont}a.
The two branches in \eqref{regBessel}  match only when $X^{\pm}(0)=0$ \ie, for $k=k_{2\ell+1}$.

In the symmetric fitting in FIG.\ref{discont}b we have  $X^{-}(0)=X^{+}(0)$, however the slopes do not
match, $(X^{-})'(0)\neq (X^{+})'(0)$, except for $k=k_{2\ell}$ when both sides vanish. 

A way to understand the quantization $k_{crit}=k_m$ of the amplitude for DM is to write for $k \geq 1$ the left profile (in blue) as
$k\,e^{U} = e^{\ln k + U}\,.$
Then increasing the amplitude can be viewed as \emph{translating the trajectory along the $U$ axis}, as shown in FIG.\ref{Zhaoshift}.
Start with  $k=1$ for which the trajectory $X_-(U)$ has no zero in $\cI_{-}$, FIG.\ref{Zhaoshift}a.  Increasing  $k$ shifts the trajectory to the left and for $k = k_1 = 2.4\dots$ the previously unphysical zero of $X^-(U)$ reaches the origin, FIG.\ref{Zhaoshift}b.
Intuitively, increasing the amplitude amounts to pulling the branches in FIG.\ref{discont} apart until they fit smoothly.
In $\cI_+$ we  glue to it the trajectory  antisymmetrically,
\beq
X_{1}^{+}(U)=-X_{1}^{-}(-U)=-J_0\big(e^{\ln k_1-U})
\label{oddX+}
\eeq
to recover Fig.\ref{modd}a.

Increasing $k$ further we arrive to the next critical value $k_2=3.8\dots $, etc, for which the tangent is zero, see FIG.\ref{Zhaoshift}. Joining the bits of trajectories symmetrically,
\beq
X_{2}^{+}(U)=+X_{2}^{-}(-U)\,,
\label{evenX+}
\eeq
we get the $m=2$ wave shown in Fig.\ref{meven}a, etc.
 In conclusion, the left and right branches
can be glued smoothly for integer half-wave number $m$,
\beq
X_{m}^{+}(U)=(-1)^{m}X_{m}^{-}(-U)\,.
\label{Xm+-}
\eeq

\begin{figure}[h]
\includegraphics[scale=.31]{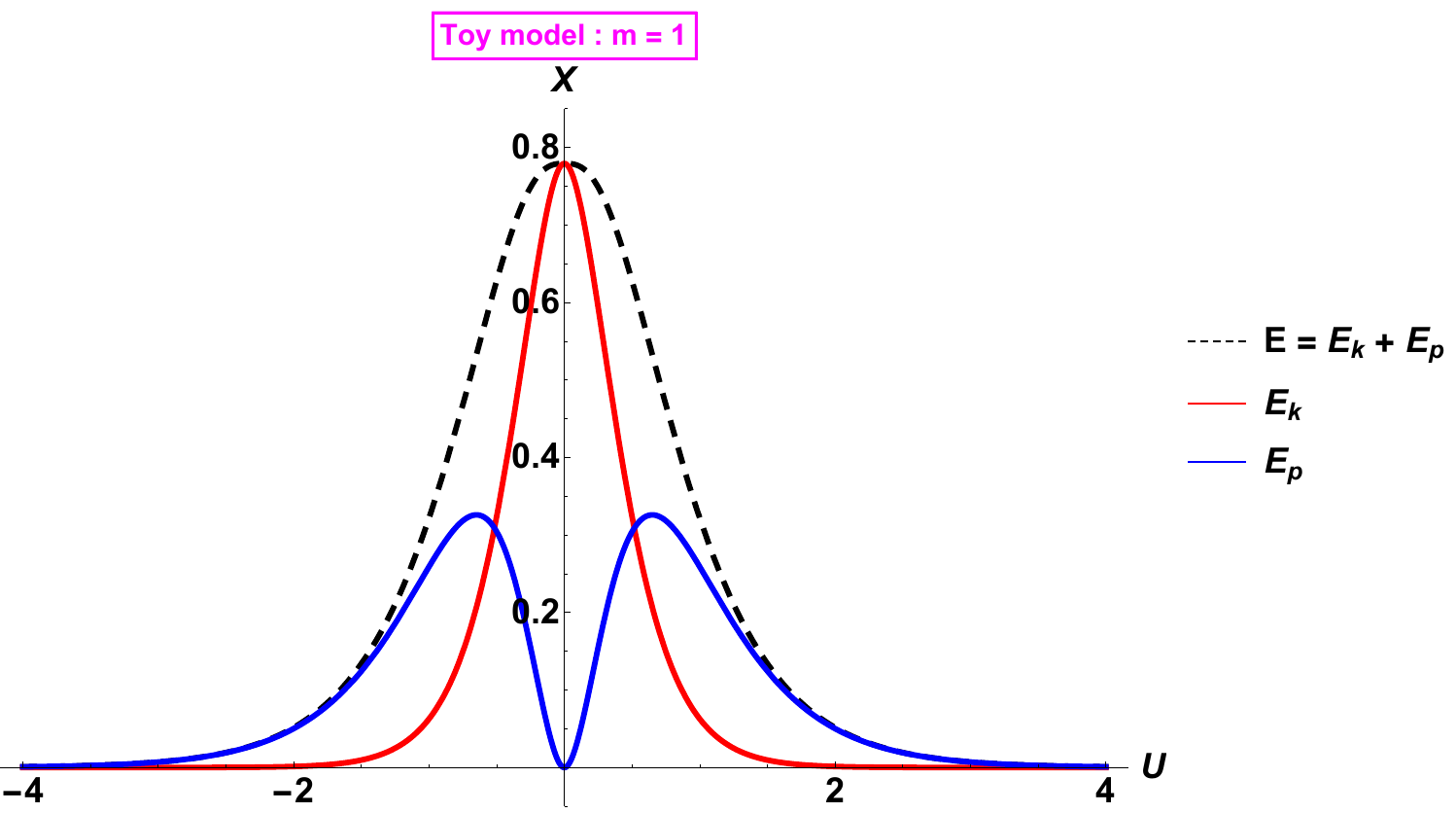}\hskip-1mm
\includegraphics[scale=.31]{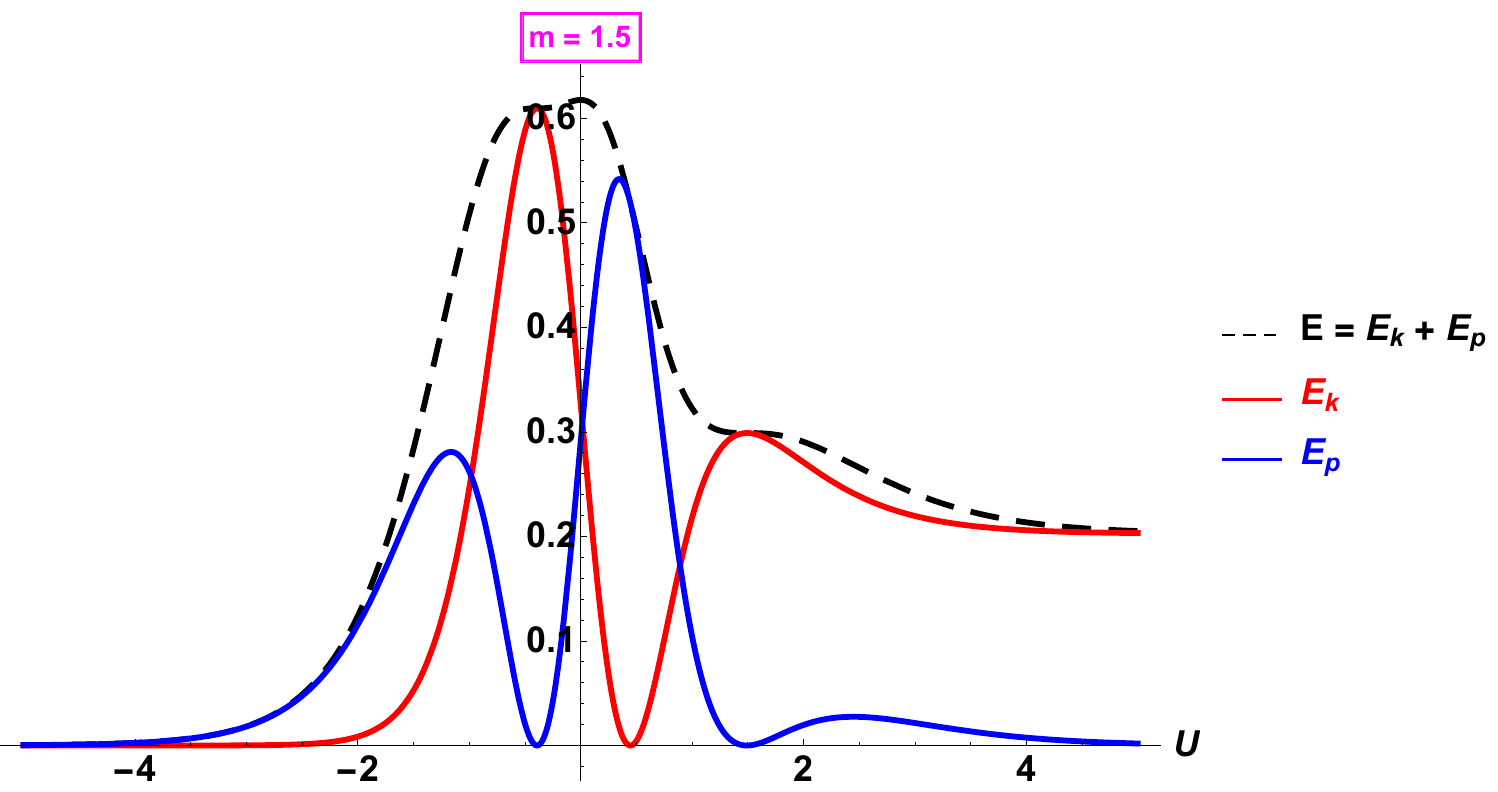}

\vskip-2mm
\hskip-18mm(a)\hskip72mm (b)
\vskip-3mm
\caption{\textit{\small For the Gaussian toy model \eqref{extoy}  the particle has  zero total energy balance for DM and positive for VM. }
\label{Ebalance}
}
\end{figure}

We conclude this section by studying the (transverse) energy  balance \cite{ZEH-PR,Carneiro}. For DM parameters total  change is \emph{zero}, as seen in FIG.\ref{Ebalance}a.
Off the critical amplitude we have (VM)~: the outgoing velocity does not vanish implying increasing energy,  consistently with FIGs. \#15 and \#6 in \cite{DM-2}. For comparison, analytic formulas can be obtained for full PT,
\besub
\begin{align}
&\hskip-6mm X_{m=1}(U)= \tanh (U)  \;
&E_{m=1}
= \frac{1}{2} \cosh(2U) \,{\rm sech}(U)^4
\label{m1PTenergy}
\\
&\hskip-6mm X_{m=2}(U)= \frac{1}{2}(3\tanh(U)^2-1)
& \;E_{m=2}
= \frac{3}{8}\big(3-2\cosh(2U) + \cosh(4U)\big)\,{\rm sech}(U)^6
\label{m2PTenergy}
\end{align}
\esub

\section{Solutions of the Sturm-Liouville eqn.}\label{SLSec}

Let us consider the solutions of the \StL eqn   \eqref{geoX2},
\beq
\dfrac {d^2\! P}{dU^2} + \cA(U)  P = 0\,.
\label{SLeq}
\eeq
For arbitrary  $k$
the general  solution with initial conditions \eqref{initcond} is,
\begin{equation}
P(U) =
\left\{
\begin{array}{cll}
aJ_{0}\left( ke^{-U}\right) +bY_{0}\left( ke^{-U}\right) &\;\text{in}\;\;&\cI_{+}
\\[4pt]
J_{0}\left( ke^{U}\right) ,~~~~&\;\text{in}\;\;&\cI_{-}%
\end{array}%
\right.\;,
\label{aJbY}
\end{equation}
where the coefficients are
\begin{equation}
a=1-\pi kY_{0}(k) J_{1}(k)=-1-\pi kY_{1}(k)J_{0}(k)\aand
b=\pi kJ_{0} (k) J_{1}(k) \,.
\end{equation}
\goodbreak
\begin{figure}[h]
\includegraphics[scale=.7]{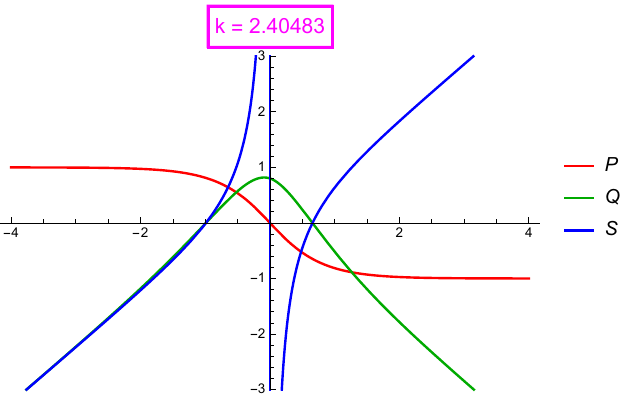}\quad\;
\includegraphics[scale=.7]{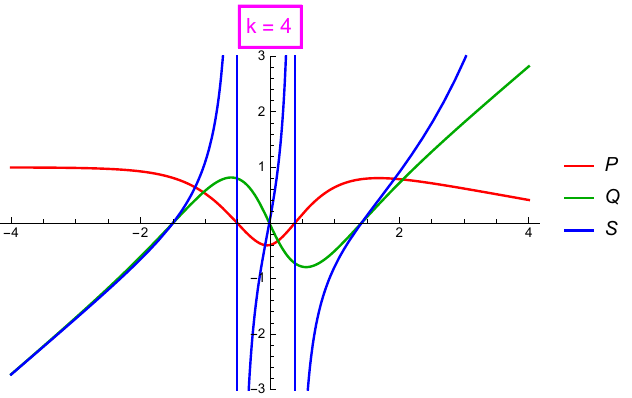}
\\
\vskip-2mm\hskip-11mm(a)\hskip76.5mm (b)
\vskip-3mm\caption{\textit{\small
The \StL solutions \red{${\bf P}$},
the non-DM 2nd solution \dgreen{${\bf Q=PS}$}
and the \blue{{\bf Souriau matrix} ${\bf S}$}
 shown here for the toy model eqn \eqref{extoy} with  (a) DM amplitude $k=k_{crit}$
 and  (b) VM amplitude $k=4$.
}
\label{ToySm1m2} }
\end{figure}

Skipping details, we mention that the 
DM strajectories $X(U) \propto P(U)$ are recovered for $k=k_{crit}$ determined by either $J_0(k) = 0$ or $J_1(k)=0$ for which $a=\pm 1$ and $b=0$\,. 

A second independent solution of (\ref{SLeq}) is,
\beq
 Q(U) =  P(U) S(U)\,,
\label{Smatrix}
\eeq
where $S(U) = \displaystyle\int^{U}_{U_0}\!\!{dv}{P^{-2}(v)}$ is
the Souriau matrix \cite{Sou73,Carroll4GW,GlobalCarroll}, which generalizes  the scalar expression considered by Arnold  in the isotropic case \cite{Arnold,SilaPT}.
All ingredients are shown in FIG.\ref{ToySm1m2}.

\section{Carroll symmetry}\label{CarrolSec}

We recall some facts. For details see \cite{GlobalCarroll}.
$P$ and $Q$ introduced above span \emph{Carroll symmetry}, generated by \cite{LeblondCar,Sou73,Carroll4GW},
\begin{equation}
h \frac{\partial}{\partial V} +
\underbrace{c \left(P \frac{\partial}{\partial X} - P' X \frac{\partial}{\partial V}\right)}_{translations}
+ \underbrace{b \left(Q  \frac{\partial}{\partial X} - Q' X  \frac{\partial}{\partial V}\right)}_{Carroll\, boosts}\,.
\label{CarrollBrink}
\end{equation}
Note here the r\^ole of the second non-DM charge $Q=PS$ in \eqref{QPSdef} as a Carroll boost generator.
The associated conserved quantities are,
\beq
\bp_0= P\bX'-\big(P\big)' \bX
 \;\aand\;
\bk_0= -Q\bX' + \big(Q\big)'\bX\,.
\label{Cconsquant}
\eeq
interpreted as the conserved linear and boost momentum, respectively.
$\p/\p_V $ generates the  mass of the underlying non-relativistic model.
Conversely, these Noether  quantities
 determine the geodesics \cite{GlobalCarroll}
\footnote{The vertical component $V$ has a more complicated variation in the Wavezone, see eqn. \#(V.5) and FIG. \# 12 of ref.\cite{DM-1}, and FIG.4 of \cite{GlobalCarroll}. Outside the Wavezone  the additional terms fall off leaving is with $V=V_0=\const$, though.},
\beq
\bX(U) = P(U)\, \bk_0 + Q(U) \,\bp_0 \,.
\label{XPQ}
\eeq

Eqn. \eqref{XPQ}  has a remarkable structure.
Firstly, the initial conditions \eqref{initcond} for $P$ imply that the conserved momentum is zero \cite{DM-1, GlobalCarroll},
\beq
\bp_0=0\,,
\label{zeromom}
\eeq
leaving us with the $P$-term alone.
DM is obtained in particular for critical amplitudes $k_{crit}$. For $k\neq k_{crit}$ we get VM as seen in FIG.\ref{ToySm1m2}a-b.

\section{Approximate Gaussian profiles}\label{GaussProf}
\begin{figure}[h]
\includegraphics[scale=.84]{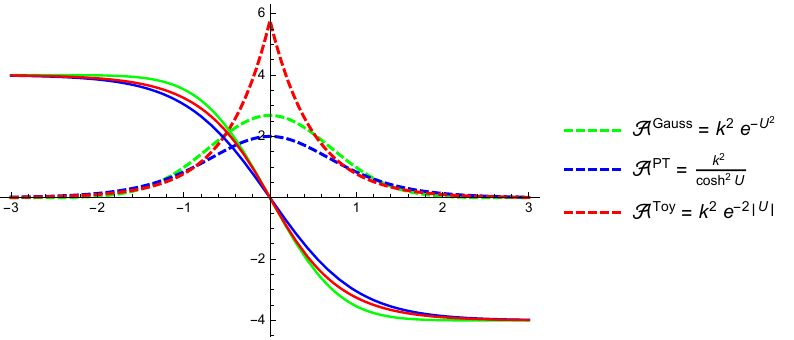}

\vskip-5mm\caption{\textit{\small
For suitably chosen parameters, the Gaussian profile and trajectory can also be approximated by that of
 \PT\!, where the coefficients are $k_{Gauss}=1.638,~k_{PT}=\sqrt{2},~k_{Toy}=2.4$. }
\label{ToyGaussprof} }
\end{figure}

Our approximation method is not limited to \PT. One can consider, for example, a  Gaussian profile  \cite{Chakra,DM-1}
\begin{equation}
\cA^{Gauss}(U) \; \propto\,e^{-U^{2}}\,,
\label{Gaussprof}
\end{equation}
whose shape can be made similar to \PT\! by fine-tuning, as seen in  FIG.\#6 of ref. \cite{DM-1}.
 For large $|U|$, the profile can again be approximated by that of the \underline{same} toy profile as for \PT, FIG.\ref{ToyGaussprof}. Gaussian and \PT geodesics  are similar to those  in \cite{DM-1}
despite their different critical amplitudes~:
$k_1=\sqrt{2}, \, k_2= \sqrt{6}$, etc for PT, and $k_1= 1.638\,, k_2=2.941\, \dots \,,$ etc for Gauss \cite{DM-1}.

\goodbreak

\begin{figure}[h]
\includegraphics[scale=.53]{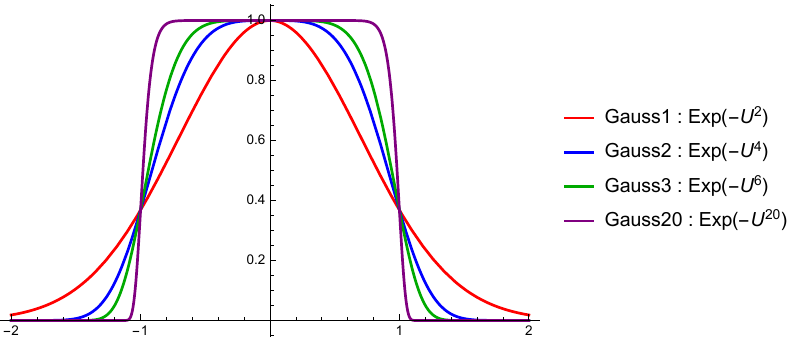}
\vskip-4mm
\caption{\textit{\small Expanded  Gaussian
$\exp[-\,U^{2q}]$, shown for  $q=1, 2, 3, 10$.
}
\label{pGaussFIG}
}
\end{figure}

\begin{figure}[h]\hskip-4mm
\includegraphics[scale=.18]{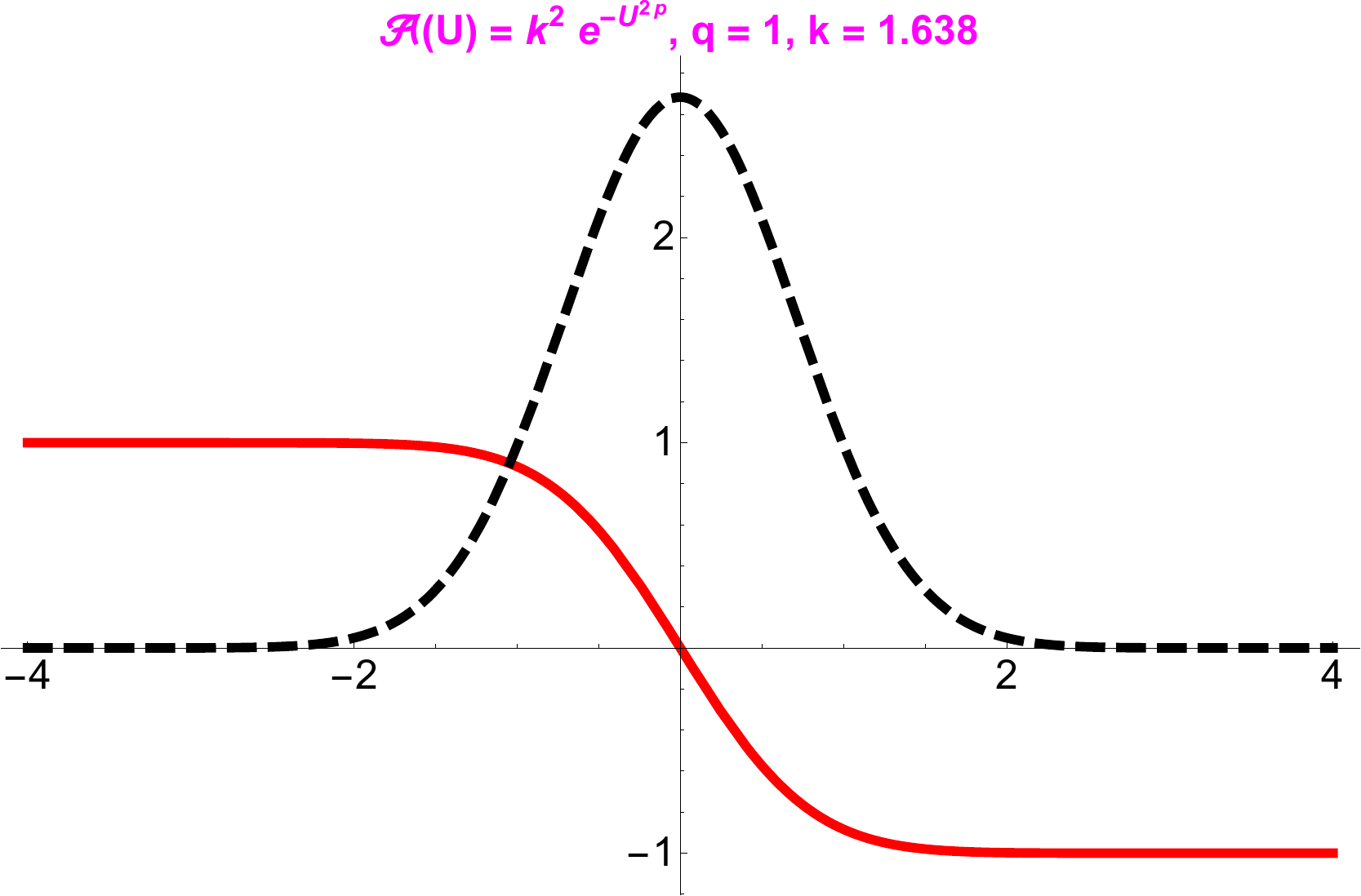}\hskip 0mm
\includegraphics[scale=.18]{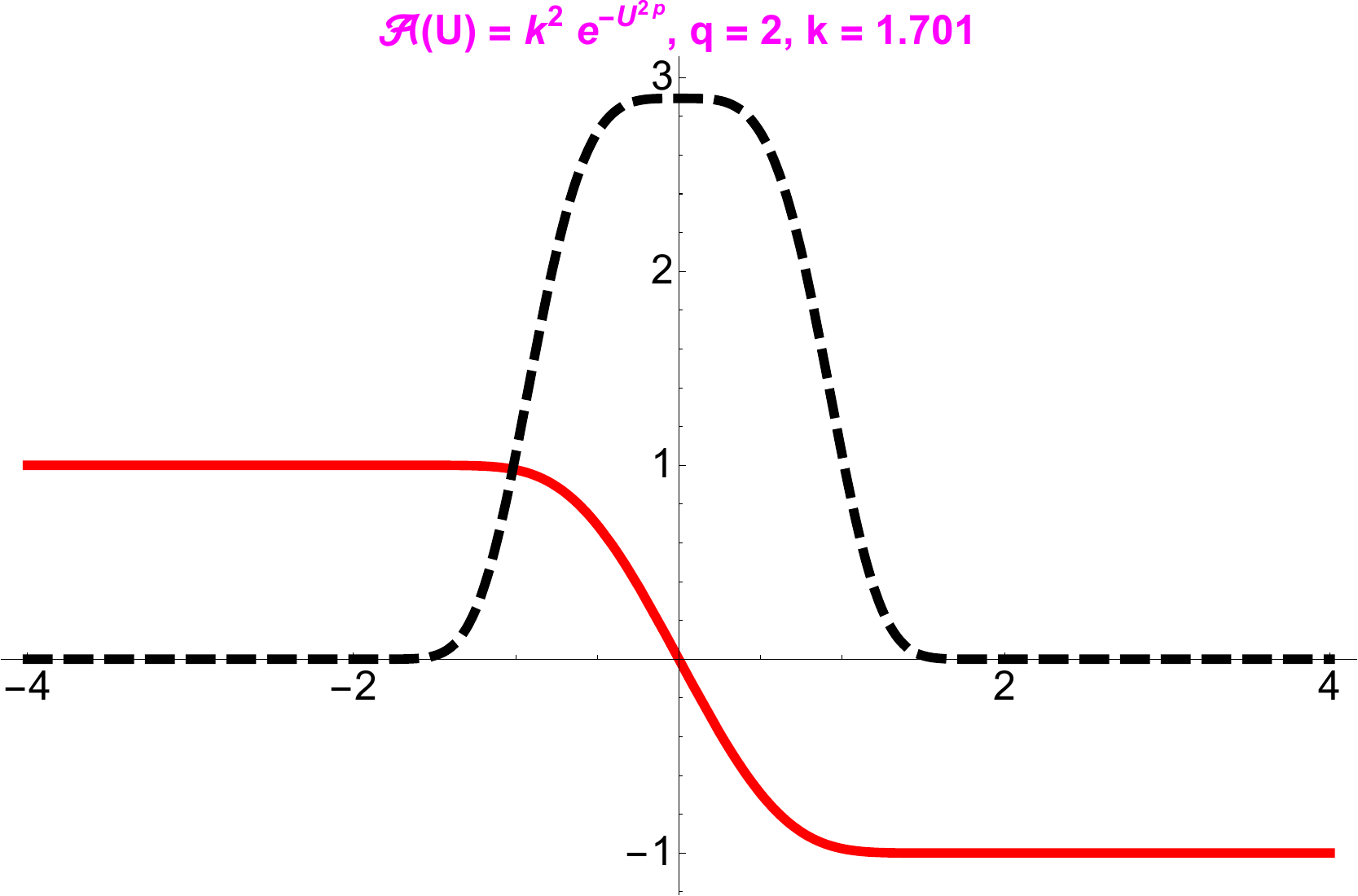}\hskip 0mm
\includegraphics[scale=.18]{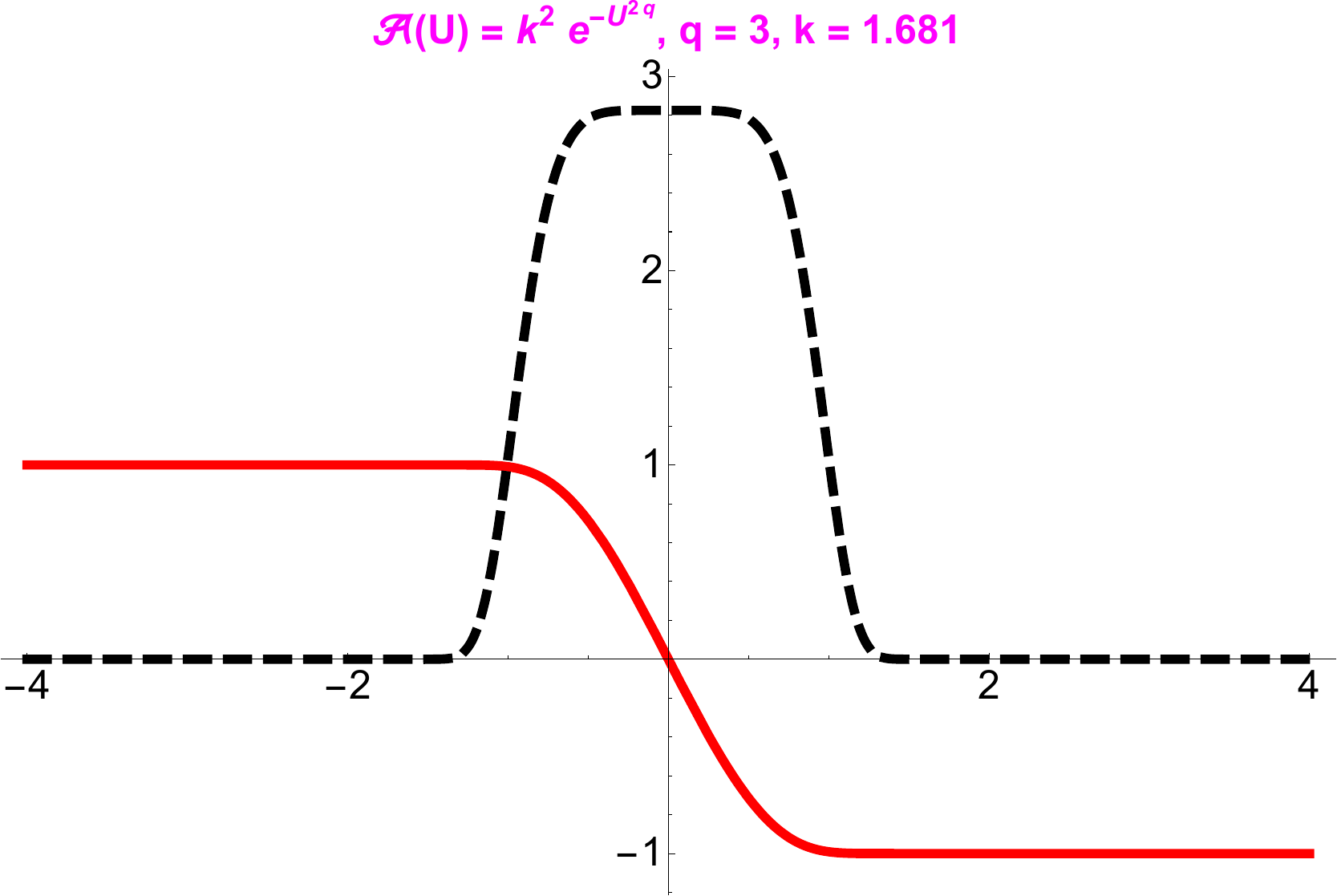}
\vskip-5mm
\caption{\textit{\small DM trajectories (red solid line) for the expanded  Gaussian profile \eqref{plimit} (dashed line),
shown for $q=1, 2, 3$.}
\label{pGaussTraj}
}
\end{figure}

We can also consider the \emph{expanded Gaussian} profile,
\beq
\cA^{exp} = k^2\,e^{- U^{2q}} \with q = 1, 2, \dots ,
\label{plimit}
\eeq
shown in FIG.\ref{pGaussFIG}. DM trajectories found by fine-tuning the amplitude, are depicted in  FIG.\ref{pGaussTraj}.
%

The $q\to\infty$ limit of  \eqref{plimit} yields  a \emph{square profile} considered in \cite{Kar3,Benin}. Both the profile and the trajectories tend to those of
\beq
{\cal A}\left( U\right) = \left\{
\begin{array}{c}
0\text{ \ \ \ \ \ \ \ \ }U\,<-1
\\[-4pt]
k^{2}\text{ \ \ }-1<U<1
\\[-4pt]
0\,\text{ \ \ \ \ \ \ \ \ \ }U\,>1\,%
\end{array}\right. \;. 
\label{1squarepot}
\eeq
Then the \StL eqns  \eqref{SLeq} yield  DM trajectories $X_m(U)=P_m(U)X_0$ with
\beq
P_m(U)=
\left\{
\begin{array}{cc}
1\; \text{ \ \ }&U<-1 \\
\cos \left[k_m(U+1)\right]  \text{ \ }&-1<U<1 \\
\:\left(-1\right)^{m} \text{ \ \ \ }&U>1 %
\end{array}%
\right.,
\quad
k_m =\frac{\pi}{2}\,m\,,~~~m=1,2,3,...  
\label{squaretraj}
\eeq

\begin{figure}[h]
\includegraphics[scale=.7]{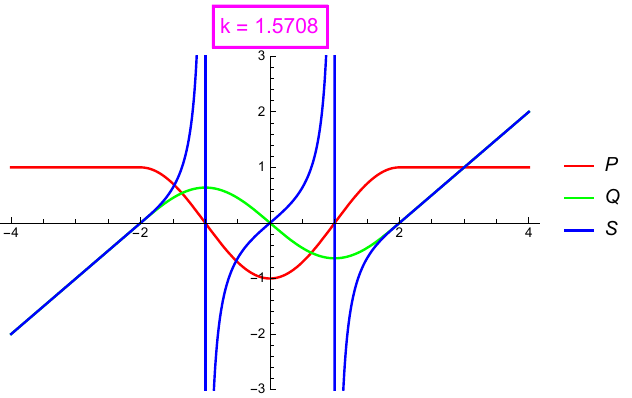}\;
\includegraphics[scale=.7]{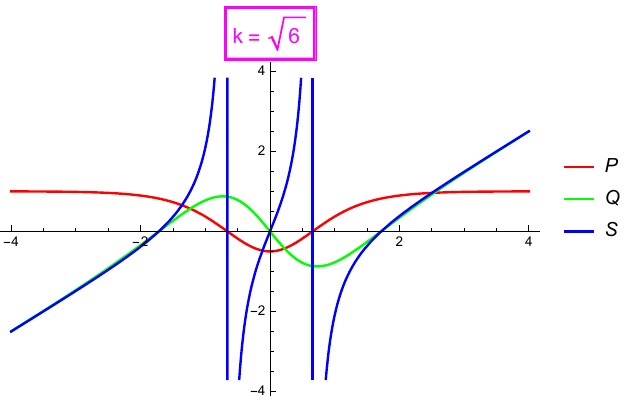}
\\
\vskip-3mm\hskip-11mm(a)\hskip72.5mm (b)
\vskip-4mm
\caption{\textit{\small The \red{\bf trajectories}, \blue{\bf   Souriau matrices} and 2nd \dgreen{\bf \StL solutions} of (a)   the square approximation \eqref{1squarepot}  show  strong similarity with those (b) for the \PT-based approximate model \eqref{extoy}, as illustrated for wave number $m=2$. }
\label{square-m2}
}
\end{figure}

\begin{figure}[h]\hskip-4mm
\includegraphics[scale=.33]{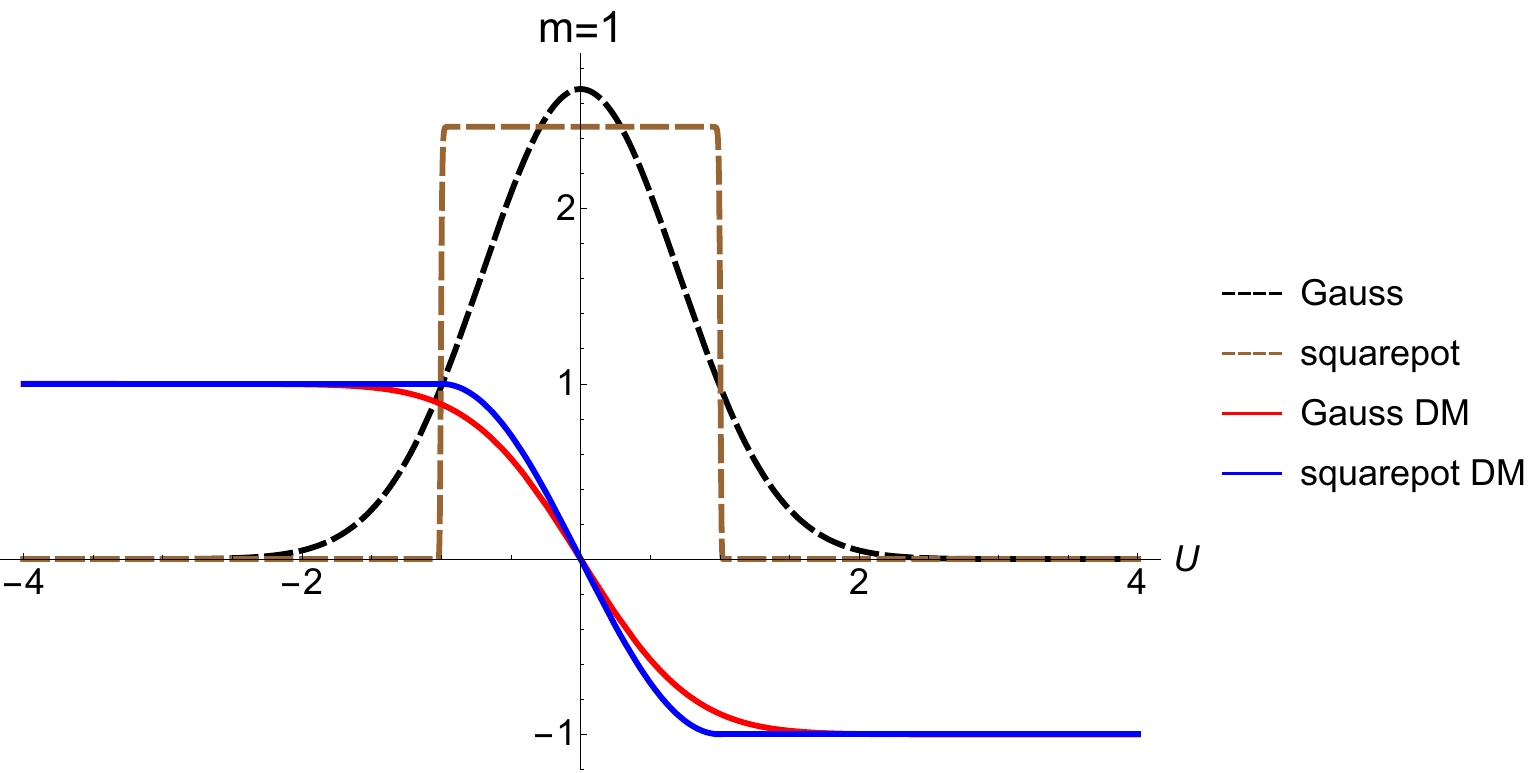}\hskip-5mm
\includegraphics[scale=.33]{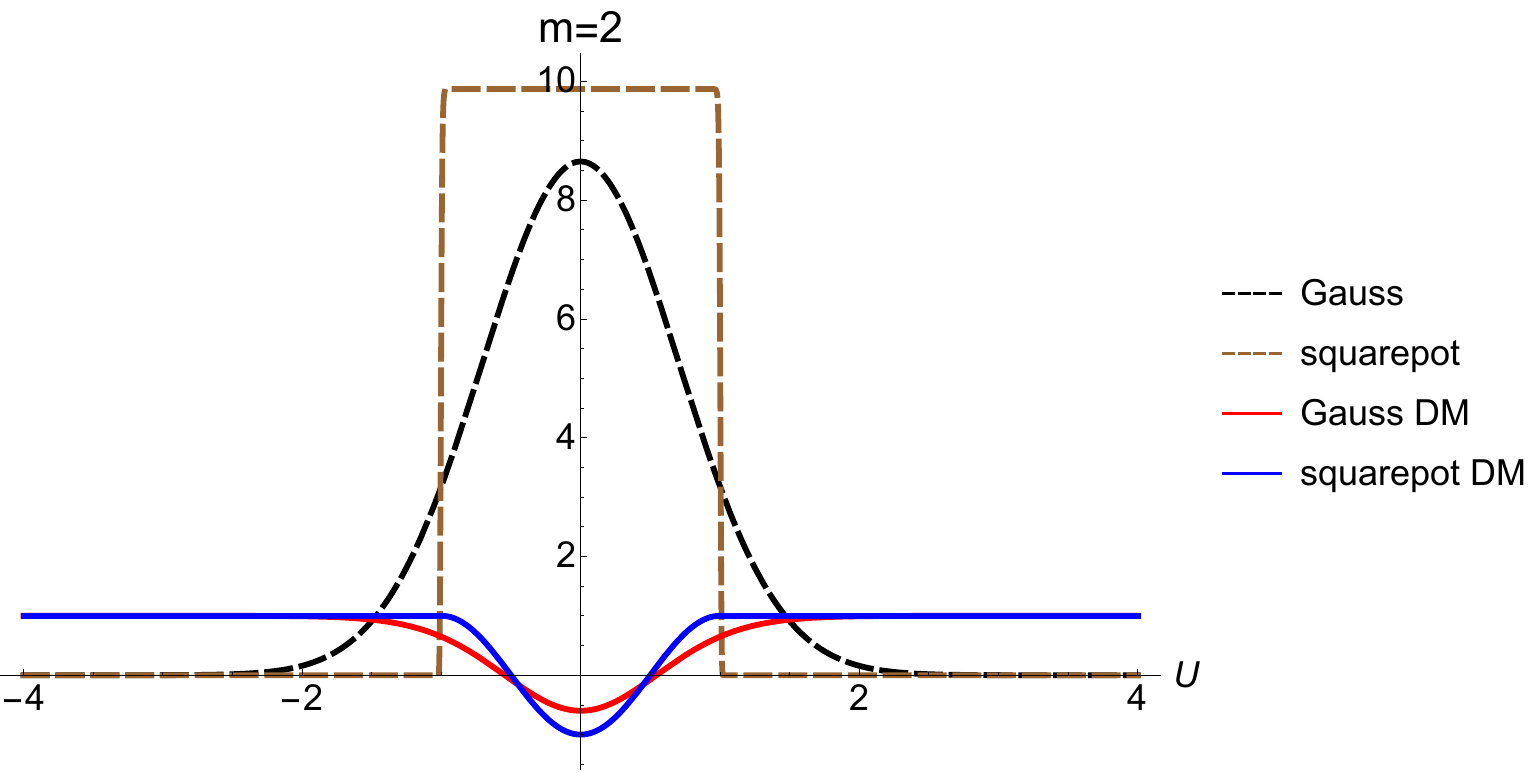}
\\
\vskip-3mm\hskip-25mm(a)\hskip76mm (b)
\vskip-4mm
\caption{\textit{\small Different wave profiles may yield similar trajectories, as illustrated by the Gaussian \eqref{Gaussprof} vs the square potential \eqref{1squarepot}.}
\label{GaussToy} }
\end{figure}

The analytic construction $P \to S \to Q=PS$ mentionned above then yields the 2nd non-DM solution shown in FIG.\ref{square-m2},
\beq
Q_m\left(U\right) =\left\{
\begin{array}{lc}
U+1,\text{ \ \ }&U<-1
\\
\frac{1}{k_m}\sin \left(k_m\left(U+1\right) \right), \text{ \ }&-1<U<1
\\[3pt]
\left( -1\right)^{m}\left(U-1\right) \text{ \ }&U>1 %
\end{array}%
\right. \quad \;
\eeq
Note the manifest similarity between
FIG.\ref{square-m2} and FIG.\ref{ToySm1m2}
 despite their different profiles, shown in FIG.\ref{GaussToy}. Further deformations will be considered in \cite{ZEH-PR}.
 
{Square profiles could also be combined antisymmetrically \cite{Benin}, yielding a double-square approximation for the flyby profile \cite{ZelPol,DM-2},
\begin{equation}
\cA \equiv \cA^{G}=
\frac{\;\;\,d}{dU}\left(\frac{k}{\sqrt{\pi}}%
e^{-U^{2}}\right)\,
\label{flybyprof}
\end{equation}%
 The DM trajectories are depicted in FIG.\ref{D2dsquareTraj}.

\begin{figure}[h]
\includegraphics[scale=.22]{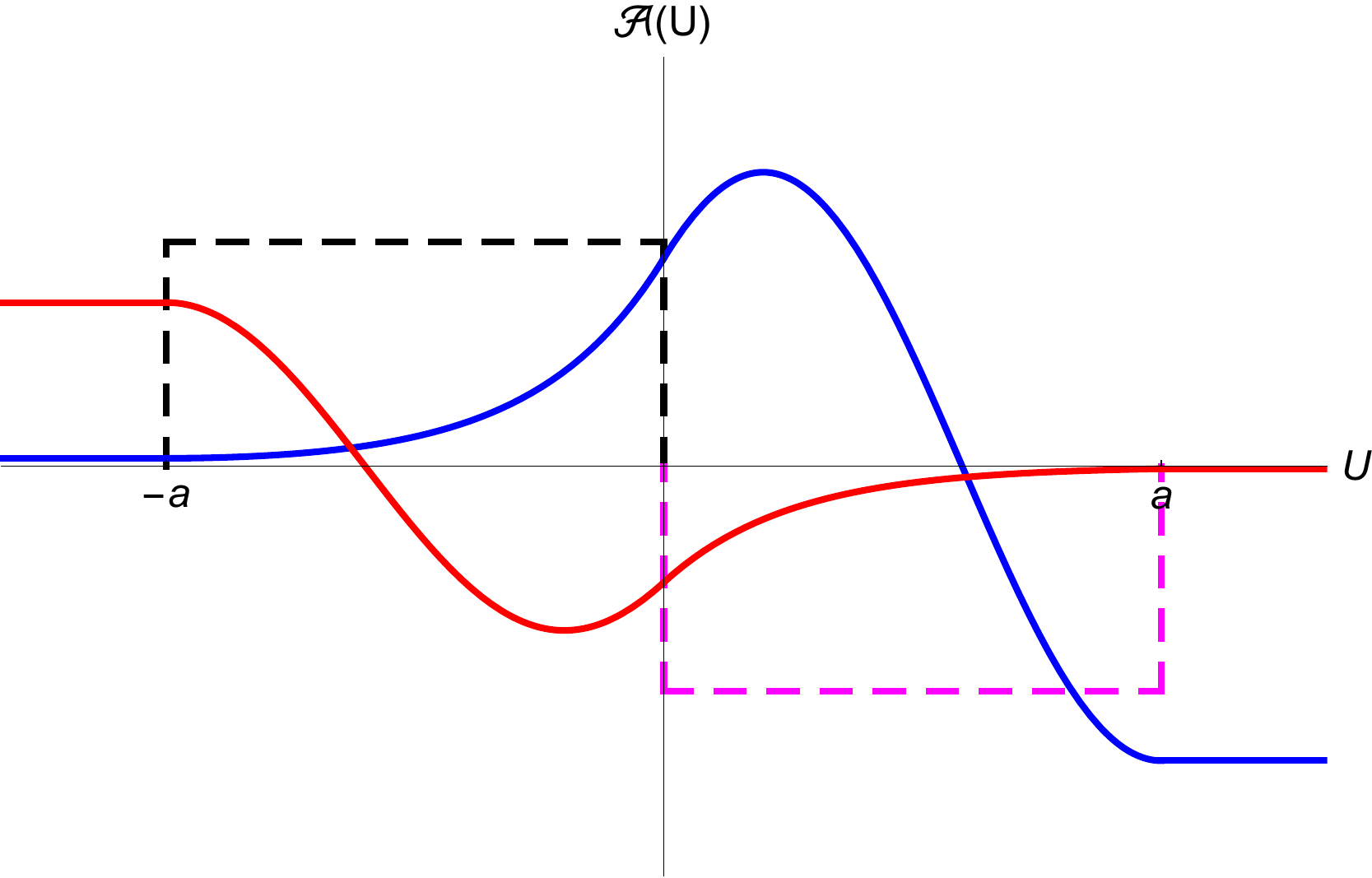} \qquad
\includegraphics[scale=.22]{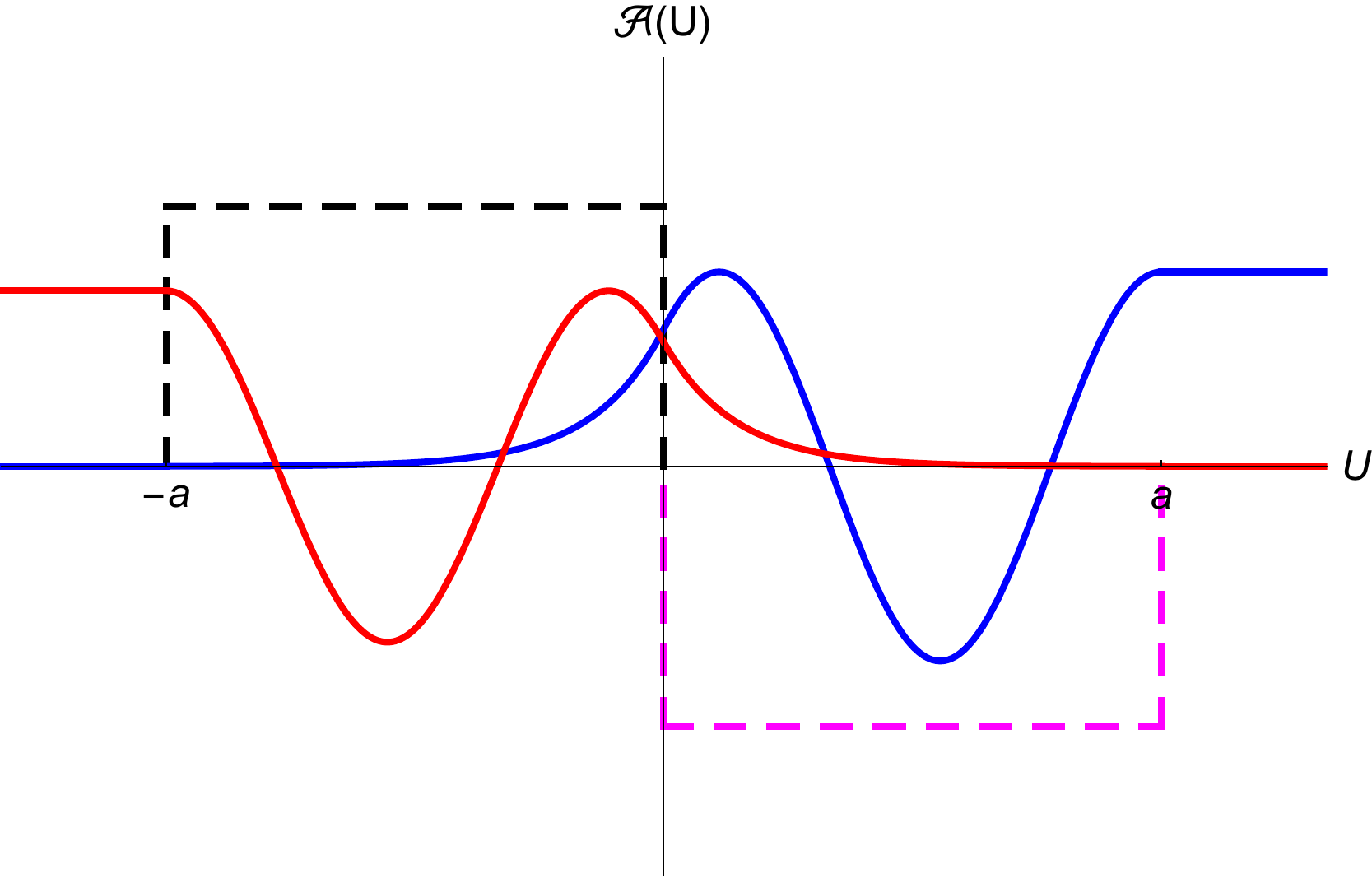}
\vskip-3mm
\hskip-2mm(a)\hskip67mm(b)

\vskip-3mm
\caption{\textit{\red{$D=2$} double-square approximation of the flyby profile \eqref{flybyprof} with wave numbers ${\bf m=1}$ and ${\bf m=2}$. The parity-dependent $U$-inversion antisymmetry/symmetry is manifest for both the \red{red} and \blue{blue} components $X^{\pm}(U)$}.
\label{D2dsquareTraj}
}
\end{figure}
The square approximation can also be extended to other singular profiles  \cite{APrencel-1,APrencel-2,Ilderton,ZCEH,Benin}.
\goodbreak

\section{Conclusion and outlook}\label{Concl}

Approximate profiles make calculations much simpler, while capturing various properties of the exact ones.  Their geodesics can be obtained by gluing together 
 two branches before and after the singularity of the profile, \eqref{regBessel}. Smooth DM geodesics are obtained for ``magic'' values $k_{crit}$ of the wave amplitude,
 which correspond  to having an integer number of half-waves in the Wavezone. Analytic calculations indicate that the \DM effect occurs  regardless of the details in the wave zone as long as the tail decays rapidly or even disappears completely.

The trajectories are consistent with supersymmetry \cite{SUSY}. Further developments \cite{Carneiro,A-Kar, Muscolino,Subir} will be considered in \cite{ZEH-PR}.

The   {Sturm-Liouville equation}
\eqref{SLeq} plays  multiple roles in our investigations \cite{Heart,EZHRev}~:
its solutions  (i) yield the particle trajectories, \eqref{XPQ} \cite{Carroll4GW};
(ii) provide us with the Carroll symmetries and conserved quantities, \eqref{CarrollBrink}--\eqref{Cconsquant}
\cite{Carroll4GW,GlobalCarroll}.

\begin{acknowledgments} \vskip-3mm
The authors are indebted to J. Balog for discussions and to Z. Silagadze for having called our attention at the approch of Arnold \cite{Arnold}.
PMZ was partially supported by the National Natural Science Foundation of China (Grant No. 12375084).
ME is supported by The Scientific and Technological Research Council of Turkey (T\"UBITAK) under grant number 125F021.
\end{acknowledgments}
\goodbreak

\end{document}